\documentclass[aps,pra,10pt,superscriptaddress,preprint]{revtex4-1}
\usepackage{amssymb}
\usepackage{amsfonts}
\usepackage{amsmath}
\usepackage{graphicx}
\usepackage{dcolumn}
\usepackage{bm}
\usepackage{subfigure}
\usepackage{color}
\usepackage{ifpdf}
\usepackage{epstopdf}
\usepackage[pdfstartview=FitH, bookmarks=true, bookmarksopen=true, breaklinks=true, colorlinks=true, linkcolor=blue, anchorcolor=blue, citecolor=blue, filecolor=blue,  menucolor=blue, urlcolor=blue]{hyperref}

\def \be{\begin{equation}}
\def \ee{\end{equation}}
\def \bs{\begin{split}}
\def \es{\end{split}}
\def \ba{\begin{array}}
\def \ea{\end{array}}
\def \bea{\begin{eqnarray}}
\def \eea{\end{eqnarray}}

\begin{document}

\title{Interference effect of critical ultra-cold atomic Bose gases}
\author{Xuguang Yue}
\affiliation{Wilczek Quantum Center, Zhejiang University of Technology,
Hangzhou 310023, China}
\affiliation{Department of Applied Physics, Zhejiang University of Technology,
Hangzhou 310023, China}

\author{Shujuan Liu}
\affiliation{Wilczek Quantum Center, Zhejiang University of Technology,
Hangzhou 310023, China}
\affiliation{Department of Applied Physics, Zhejiang University of Technology,
Hangzhou 310023, China}

\author{Hongwei Xiong}
\thanks{Electronic address: xionghw@zjut.edu.cn}
\affiliation{Wilczek Quantum Center, Zhejiang University of Technology,
Hangzhou 310023, China}
\affiliation{Department of Applied Physics, Zhejiang University of Technology,
Hangzhou 310023, China}

\begin{abstract}
For ultra-cold atomic gases close to the critical temperature, there is a divergent correlation behavior within the critical regime. This divergent correlation behavior is the cornerstone of the universal behavior within the critical regime, e.g. the universal critical exponent for the same class with very different physical systems. It is still quite challenging to observe this divergent correlation behavior in experiments with ultra-cold atomic gases. Here we consider theoretically the interference effect of the critical atomic Bose gas by a Kapitza-Dirac scattering. We find that the Kapitza-Dirac scattering has the merit of enhancing the interference effect in the observation of the correlation behavior. This provides a potential method to study the critical behavior of ultra-cold Bose gases. A simple rule is found by numerical simulations to get the critical exponent and correlation amplitude ratio from the interference fringes after the Kapitza-Dirac scattering.
\end{abstract}
\maketitle

\section{Introduction}
Phase transitions are common in nature, ranging from the cooling of the early universe, the freezing of water into ice, to the $\lambda$-transition of liquid helium~\cite{Zinn-Justin2013, Navon2015Sci, Zurek1996PhysRep}. A typical phase transition can usually be described by the order parameter. Near the transitions, the fluctuations of the order parameter dominate, leading to divergent correlation, with the behavior of the system governed by the correlation length $\xi$. Consequently, the macroscopic quantities of the system show a universal scaling behavior characterized by the critical exponents, and the phase transitions can be classified into a small number of universality classes, depending on the generic properties of the system such as dimensionality, symmetries, and range of interactions~\cite{Sachdev2011}.

The realization of Bose-Einstein condensate (BEC) in dilute atomic gases is one of the most remarkable achievements in the observation of phase transitions of novel physical systems. The transition from a thermal gas to a condensate represents a prototype of one of the universality classes, i.e. the so-called three-dimensional XY model with interaction~\cite{Zinn-Justin2002}.
Because of the exquisite control over various system parameters, the cold atomic gases have become unique platforms suitable to study the critical phenomenon~\cite{Bloch2008RMP, Stoof2008}. In a landmark experiment~\cite{Donner2007Science}, the critical behavior in the correlation length was studied by detecting the interference of two released atomic clouds.
Recently, it was demonstrated experimentally that this critical phase transition could also be revealed using a Talbot-Lau interferometer~\cite{Xiong2013LPL}.

In the critical regime, the most powerful method to demonstrate the spatial coherence is by interference. In the pioneering work in Ref.~\cite{Donner2007Science}, the spatial coherence of the Bose gas within the critical regime is studied by the releasing and subsequent interference of two clouds in the Bose gas, which is similar to double-slit interference. In Ref.~\cite{Chomaz2015nc}, the phase coherence of a quasi two-dimensional Bose gas is probed by the interference between two independent Bose gases. In the present work, we consider the application of Kapitza-Dirac (KD) scattering~\cite{Cronin2009RMP, Yue2013PRA} to probe the spatial coherence of the Bose gas in the critical regime. For ultra-cold Bose gases, the KD scattering is realized by a pulsed periodic potential, which plays a role of multiple-split diffraction, compared with the double-split interference method in Ref.~\cite{Donner2007Science}.
From the dynamics of the one-body density matrix after a short sudden change to the Hamiltonian of the system~\cite{Polkovnikov2011RMP} because of the KD scattering, we predict clear interference pattern within the critical regime. We find that the critical correlation can be revealed from the momentum distribution.
The critical temperature can be identified by measuring the visibility of the side peaks in the momentum distribution. Furthermore, a simple rule is found to extract the critical exponent and amplitude ratio of the correlation length from the dependence of the visibility on the relative temperature. Compared with the standard time-of-flight (TOF) method, measuring the visibility requires quite low signal to noise ratio (SNR). Our method provides an alternative way to study the critical behavior.

The manuscript is organized as follows. In Sec.~\ref{Sec:OBDM}, we briefly introduce the one-body density matrix and its relation with the correlation function. In Sec.~\ref{Sec:CorrInCriRegime}, the dynamics of the one-body density matrix is expressed with the propagator method, which shows that the critical correlation can be revealed from the momentum distribution. In Sec.~\ref{Sec:KDS}, the general method is applied to the KD scattering of an ultra-cold Bose gas within the critical regime. Based on our theoretical simulation, we find a simple relation to extract the critical behavior. A simple summary and discussion is given in Sec.~\ref{Sec:Disc}.

\section{One-body density matrix and critical correlation}

\label{Sec:OBDM}

The onset of a BEC can be defined through the emergence of the off-diagonal long range order of the one-body density matrix, irrespective of the presence or absence of interactions and external trap~\cite{Penrose1956PR, Pitaevskii2003, Ueda2010}. Under this consideration, we introduce the one-body density matrix
\begin{equation}
n^{(1)}\left(\mathbf{r}_{1},\mathbf{r}_{2}\right)=\left\langle \hat{\Psi}^{\dag} \left(\mathbf{r}_{1}\right)\hat{\Psi}\left(\mathbf{r}_{2}\right)\right\rangle\,,
\end{equation}
where $\hat{\Psi}\left(\mathbf{r}\right)$ ($\hat{\Psi}^{\dag}\left(\mathbf{r}\right) $) is the Bose field operator annihilating (creating) a particle at position $\mathbf{r}$, obeying the bosonic commutation relation $[\hat{\Psi}\left(\mathbf{r}\right), \hat{\Psi}^{\dag} \left(\mathbf{r}^{\prime}\right)] =\delta(\mathbf{r} -\mathbf{r}^{\prime})$. $\langle \cdot \rangle$ considered in this work means statistical ensemble average with the quantum average for a pure state as a special case. It is apparent that the spatial correlation between different $\mathbf{r}_1$ and $\mathbf{r}_2$ can be known directly if we have the information of the one-body density matrix. For the special case $\mathbf{r}_{1} =\mathbf{r}_{2}$, the one-body density matrix reduces to the average density distribution $n\left(\mathbf{r}\right)$, whose integral over the spatial coordinate gives the total number of particles $N=\int d\mathbf{r}n\left(\mathbf{r}\right)=\int d\mathbf{r}n^{(1)}\left(\mathbf{r},\mathbf{r}\right)$.

In the critical regime of the phase transition to a BEC, almost all atoms are correlated to each other because of the presence of large fluctuations.  In this case, it is usually impossible to get the accurate and analytic expression of the many-body quantum state. However,
although the fluctuations are large and the spatial correlation is enhanced, which make the many-body quantum state extremely complex, there is a simple and universal behavior from the approximate result of the one-body density matrix based on the powerful renormalization group theory~\cite{Wilson1975RMP}.
Consequently, the information of the critical behavior can be extracted from the one-body density matrix.

As an example, we consider a uniform three-dimensional dilute Bose gases in the critical regime. The one-body density matrix can be written as
\begin{equation}
n^{(1)}\left(\mathbf{r}_{1},\mathbf{r}_{2}\right) =\sqrt{n\left(\mathbf{r}_{1}\right)} \sqrt{n\left( \mathbf{r}_{2}\right)}g^{\left(1\right)} \left(\mathbf{r}_{1},\mathbf{r}_{2}\right),
\label{eq:n1g1}
\end{equation}
where $g^{\left(1\right)}\left(\mathbf{r}_{1},\mathbf{r}_{2}\right)$ is a dimensionless function reflecting the correlation of atoms at different locations. Here $n\left(\mathbf{r}\right)$ is the density distribution in thermal equilibrium. The correlation function provides a direct information of the many-body excitations and their weight, describes complex many-body states, and gives particularly important information about phase transitions, where they exhibit characteristic scaling form~\cite{Sachdev2011}.

For a uniform and dilute three-dimensional Bose gas in the critical regime, $g^{\left(1\right)}\left(\mathbf{r}_{1},\mathbf{r}_{2}\right)=g^{\left(1\right)} \left( \mathbf{r}_{1}-\mathbf{r}_{2}\right)$ takes the following well-known expression~\cite{Huang1987}
\begin{equation}
g^{\left(1\right)}\left(\mathbf{r}_{1},\mathbf{r}_{2}\right)\propto\frac{1}{\left\vert \mathbf{r}_{1}-\mathbf{r}_{2}\right\vert}e^{-\left\vert \mathbf{r}_{1}-\mathbf{r}_{2}\right\vert/\xi}.
\label{eq:corr}
\end{equation}
Of course, this expression is invalid when $\mathbf{r}_1=\mathbf{r}_2$. In fact it should be used when $|\mathbf{r}_1-\mathbf{r}_2|>\lambda_T$ with $\lambda_T$ the thermal de Broglie wavelength. In Eq.~(\ref{eq:corr}), the correlation length is given by
\be
\xi(T)=\left\{\begin{array}{rcl}
a_{>}\left|\frac{T_c}{T-T_c}\right|^{v_{>}}, & \mbox{for} & T>T_c,\\[8pt]
a_{<}\left|\frac{T_c}{T-T_c}\right|^{v_{<}}, & \mbox{for} & T<T_c.
\end{array}\right.
\ee
Here $\nu_{>}$ and ${\nu_{<}}$ are the critical exponents above and below the critical temperature, respectively. Present theoretical models based on three-dimensional field theory, $\varepsilon$-expansion method and other numerical methods show that $\nu_{>}=\nu_{<}$~\cite{Zinn-Justin2013}. However, it is possible that below and above the critical temperature, the amplitude $a_{>}\neq a_{>}$. So far, different theoretical models give different amplitude ratio $a_{>}/a_{<}$. The three-dimensional field theory gives the theoretical value $a_{>}/a_{<}\approx 0.50$, while the $\varepsilon$-expansion method predicts $a_{>}/a_{<}\approx 0.33$~\cite{Privman1991}. As the variation in amplitude ratio is more pronounced, it provides a more stringent test of universality classes than do critical exponents~\cite{Chaikin2000}. Hence, it is appealing to consider whether the ultra-cold atomic gases provide the chance to measure this amplitude ratio.

For ultra-cold atomic gases, the information of the system for most experiments is obtained by sufficiently long free expansion, so that the measured density distribution reflects directly the momentum distribution. From the relation between the field operator $\hat{\Psi}\left(\mathbf{r}\right)$ in space representation and that in momentum representation $\hat{\Psi}\left(\mathbf{p}\right)=\left(2\pi\hbar\right)^{-3/2}\int d\mathbf{r}e^{-i\mathbf{p\cdot r}/\hbar}\hat{\Psi}\left(\mathbf{r}\right)$,
we can get the momentum distribution $n\left(\mathbf{p}\right)=\langle \hat{\Psi}^{\dag}\left( \mathbf{p}\right)\hat{\Psi}\left(\mathbf{p}\right)\rangle $ from the one-body density matrix in thermal equilibrium,
\begin{equation}
n\left(\mathbf{p}\right)=\frac{1}{\left(2\pi\hbar\right)^{3}}\int d\mathbf{r}_1d\mathbf{r}_2n^{(1)}\left( \mathbf{r}_{1}, \mathbf{r}_{2}\right)e^{i\mathbf{p\cdot (\mathbf{r}_{1}-\mathbf{r}_{2})}/\hbar}.
\label{eq:momentum}
\end{equation}
It is quite interesting to note that as $n^{(1)}\left(\mathbf{r}_{1},\mathbf{r}_{2}\right)$ comprises the information of the correlation of the system, the measurement of $n(\mathbf{p})$ has the potential to give the information of the spatial correlation. In a recent experimental work on a uniform quasi-two-dimensional Bose gas~\cite{Chomaz2015nc}, the measurement of $n(\mathbf{p})$ is used as a method to show the phase transition.

\section{The dynamics of one-body density matrix after a sudden change}
\label{Sec:CorrInCriRegime}

In this section, we study the dynamics of the one-body density matrix after a sudden change to the system, which gives us the foundation to use KD scattering to investigate the critical phenomenon in section~\ref{Sec:KDS}.

We begin with rewriting the one-body density matrix by expanding the field operator $\hat{\Psi} \left(\mathbf{r}\right)=\sum_{j}\psi_{j}\left(\mathbf{r}\right)\hat{a}_{j}$ with $\{\psi_{j}\left(\mathbf{r}\right)\}$
forming a complete orthogonal basis as
\begin{equation}
n^{(1)}\left(\mathbf{r}_{1},\mathbf{r}_{2},t_i\right)=\sum_{i,j}\psi_{i}^{\ast}\left(\mathbf{r}_{1},t_i\right) \psi_{j}\left(\mathbf{r}_{2},t_i\right)\left\langle \hat{a}_{i}^{\dag}\hat{a}_{j}\right\rangle.
\label{eq:initialOB}
\end{equation}
Here we have included the time variable explicitly. $t_i$ denotes the time before the change and the system is in equilibrium.

We now consider the dynamical evolution of the one-body density matrix after a sudden change to the system. Without loss of generality, we assume the change is applied between $t=t_{i}$ and $t=t_{f}$. We are interested in the situation where the duration of the change $\tau=t_f-t_i$ is so short that the interaction between atoms does not change $\langle \hat{a}_{i}^{\dag}\hat{a}_{j}\rangle$. Physically, this means that during a short period after the sudden change, the redistribution between different energy levels is negligible, and hence the reestablishment of thermal equilibrium can be omitted.
In this case, the dynamical evolution of the one-body density matrix can be expressed as
\be
\begin{split}
n^{(1)}\left(\mathbf{r}_{1},\mathbf{r}_{2},t\right)
=\sum_{i,j}\int d\mathbf{r}_{1}^{\prime} K^{\ast}\left(\mathbf{r}_{1},t; \mathbf{r}_ {1}^{\prime},t_{i} \right)\psi_{i}^{\ast} \left(\mathbf{r}_{1}^{\prime},t_i\right) &  \\
\times\int d\mathbf{r}_{2}^{\prime} K\left(\mathbf{r}_{2},t;\mathbf{r}_{2}^{\prime},t_{i} \right)\psi_{j}\left(\mathbf{r}_{2}^{\prime},t_i\right)\left\langle \hat{a}_{i}^{\dag}\hat{a}_{j}\right\rangle.&
\end{split}
\label{eq:evolution}
\ee
Here $K\left(\mathbf{r},t;\mathbf{r}^{\prime},t_{i}\right)$ is the propagator of the Hamiltonian $H\left(\mathbf{r},t\right)$ between $t_i$ and $t_f$, satisfying
\begin{equation}
\left[i\hbar\frac{\partial}{\partial t}-H\left(\mathbf{r},t\right)\right]K(\mathbf{r},t;\mathbf{r}^{\prime},t_{i})
=i\hbar\delta(t-t_{i})\delta(\mathbf{r}-\mathbf{r}^{\prime}).
\end{equation}
The propagator $K(\mathbf{r}, t; \mathbf{r}^{\prime}, t_{i})$ is the kernel of the differential operator in question and gives the probability amplitude of a particle to travel from one place $\mathbf{r}^{\prime}$ at time $t_{i}$ to another $\mathbf{r}$ at time $t$. It can also be written as
\be
K\left(\mathbf{r}, t; \mathbf{r}^{\prime}, t_{i}\right) =\left\langle\mathbf{r}\right\vert U\left(t, t_{i}\right)\left\vert\mathbf{r}^{\prime}\right\rangle,
\label{eq:propu}
\ee
with $U\left(t, t_{i}\right) =\exp\left[-i\int_{t_i}^{t} H\left(\mathbf{r},\tau\right)d\tau/\hbar\right]$ the unitary time-evolution operator for the system. From Eqs.~(\ref{eq:initialOB}) and (\ref{eq:evolution}), we have the following simple formula about the evolution of the one-body density matrix.
\begin{equation}
\begin{split}
n^{(1)}\left(\mathbf{r}_{1},\mathbf{r}_{2},t\right)&=\int d\mathbf{r}_{1}^{\prime}d\mathbf{r}_{2}^{\prime}K^{\ast}\left(\mathbf{r}_{1},t; \mathbf{r}_{1}^{\prime}, t_i\right)\\
&\times K\left(\mathbf{r}_{2},t; \mathbf{r}_{2}^{\prime},t_i\right) n^{(1)}\left(\mathbf{r}_{1}^{\prime},\mathbf{r}_{2}^{\prime},t_i\right).
\end{split}
\label{eq:evol}
\end{equation}

For the system initially in the critical regime, substituting Eq.~(\ref{eq:n1g1}) into the above equation, it is easy to get
\begin{equation}
\begin{split}
n^{(1)}\left(\mathbf{r}_{1},\mathbf{r}_{2},t\right) &=\int d\mathbf{r}_{1}^{\prime} d\mathbf{r}_{2}^{\prime} K^{\ast}\left(\mathbf{r}_{1},t; \mathbf{r}_{1}^{\prime},t_i\right) K\left(\mathbf{r}_{2},t;\mathbf{r}
_{2}^{\prime},t_i\right) \\
&\times\sqrt{n\left(\mathbf{r}_{1}^{\prime}\right) n\left(\mathbf{r}_{2}^{\prime}\right)} g^{\left(1\right)} \left(\mathbf{r}_{1}^{\prime},\mathbf{r}_{2}^{\prime}\right).
\end{split}
\label{eq:evolcritical}
\end{equation}
In the special case when the system is perfectly coherent, \emph{i.e.}, $g^{(1)}=1$ over the whole system, Eq.~(\ref{eq:evol}) is consistent with the solution by the Schr\"{o}dinger equation. From Eq.~(\ref{eq:evolcritical}), we also have the evolution of the density distribution
\begin{equation}
\begin{split}
n\left(\mathbf{r},t\right)&=\int d\mathbf{r}_{1}^{\prime}d\mathbf{r}_{2}^{\prime}K^{\ast}\left( \mathbf{r},t;\mathbf{r}_{1}^{\prime},t_i\right)K\left(\mathbf{r},t;\mathbf{r}_{2}^{\prime},t_i\right)\\
&\times\sqrt{n\left( \mathbf{r}_{1}^{\prime}\right)n\left(\mathbf{r}_{2}^{\prime}\right)}g^{\left(1\right)}\left( \mathbf{r}_{1}^{\prime},\mathbf{r}_{2}^{\prime}\right).
\end{split}
\label{eq:denevol}
\end{equation}
Eq.~(\ref{eq:denevol}) shows that the measurement of the density distribution $n\left(\mathbf{r},t\right)$ after the change of the Hamiltonian can also give us the information of the spatial correlation of the system. Note that the measurement of the density distribution in thermal equilibrium at $t=t_i$ is not likely to give this information.

We have shown that the dynamics of the one-body density matrix can be used to diagnose the critical correlation. In the next section, we apply the general results obtained in this section to a special type of sudden change, namely the KD scattering of cold atoms. The goal is to investigate whether the KD scattering can be used to identify the phase transition and critical behavior.

\section{Momentum distribution and critical correlation after the Kapitza-Dirac scattering}
\label{Sec:KDS}

In this section we use the KD scattering to investigate the critical correlation of cold atoms near the phase transition to a BEC. We first calculate the propagator of the scattering process, with which we can get the momentum distribution after the scattering process according to Eq.~(\ref{eq:momentum}) and Eq.~(\ref{eq:evolcritical}). At last we demonstrates how to extract the information of the critical correlation from the visibility of the side peaks in the momentum distribution.

\begin{figure}[t]
\centering
\includegraphics[width=8cm]{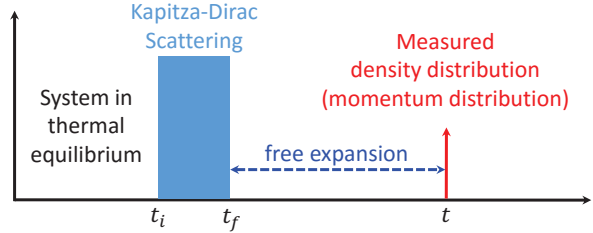}
\caption{The scheme of probing the critical behavior with the application of the KD scattering between $t_i$ and $t_f$. After sufficiently long free expansion, the measurement of the density distribution shows approximately the momentum distribution before the free expansion. This measured momentum distribution gives the information of the critical behavior which can be extracted from the interference pattern induced by the KD scattering.}
\label{fig:scheme}
\end{figure}

The scheme of probing the critical behavior with KD scattering is shown in Fig.~\ref{fig:scheme}. The KD scattering can be realized by radiating the atomic gas by a standing-wave light pulse during time $t_i$ and $t_f$. The detuning of the laser light is far from resonance so that we can ignore the effect of spontaneous emission. After the KD scattering, the system is allowed to expand freely for sufficiently long time, and we show that the measured momentum distribution can be used to extract the information of the critical correlation. For simplicity and without loss of generality, we carry out the calculations for a quasi-one-dimensional system, which physically corresponds to a three-dimensional cigar-shaped Bose gas. The Hamiltonian of the system diffracted by the standing wave light field is given by
\begin{equation}
H\left(z,t\right)=\frac{p_{z}^{2}}{2m}+U_{0}\cos^{2}\left(kz\right),
\label{eq:hamil}
\end{equation}
where the second term on the right hand side is the effective periodic potential~\cite{Denschlag2002JPB, Morsch2006RMP}, with $k=\pi/d$ the wave vector of the laser and $d$ the spatial period of the standing wave. According to Bloch's Theorem, the eigen-states of the Hamiltonian (\ref{eq:hamil}) are Bloch states  $\left\vert n,q\right\rangle $ labeled by the band index $n=0, 1, 2, \cdots$ and quasi-momentum $q$, with the corresponding eigen-energy $E_{n,q}$.

Inserting the identity $\sum_{n,q}\left\vert n,q\right\rangle \left\langle n,q\right\vert=1$ into Eq.~(\ref{eq:propu}), we get the propagator during the KD scattering process,
\be
\begin{split}
K\left(z,t;z^{\prime},t_{i}\right)=&\left\langle z\right\vert\sum_{n,q}\left\vert n,q\right\rangle \left\langle n,q\right\vert e^{-iH(z,t)(t-t_i)/\hbar}\\
&\times\sum_{n^{\prime},q^{\prime}} \left\vert n^{\prime},q^{\prime}\right\rangle \left\langle n^{\prime},q^{\prime}\right\vert z^{\prime}\rangle\\
=&\sum_{n,q}\phi_{n,q}\left(z\right) e^{-iE_{n,q}(t-t_i)/\hbar}\phi_{n,q}^{\ast}\left(z^{\prime}\right)\,,
\label{eq:propogator}
\end{split}
\ee
where
\begin{equation}
\phi_{n,q}(z)=\langle z\left\vert n,q\right\rangle =\sum_{l}c_{l}^{(n,q)}e^{i(2lk+q)z}\,,
\end{equation}
is the Bloch wave function with $c_{l}^{(n,q)}$ its Fourier expansion coefficient.

\begin{figure}[t]
\centering
\includegraphics[width=9cm]{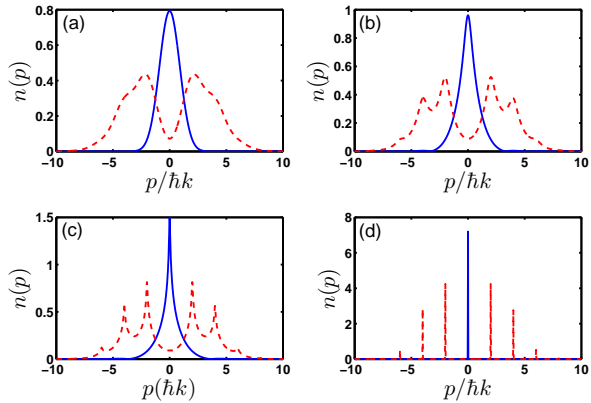}
\caption{Momentum distribution before (blue solid line) and after (red dashed line) KD scattering at different correlation length. (a) thermal atoms $T\gg T_c$; (b) and (c) $\xi=1$ and $\xi=10$ in the critical regime, respectively; (d) pure condensate at zero temperature.}
\label{fig:fig2}
\end{figure}

Assuming the initial density a Gaussian distribution $n(z)=\exp(-z^2/\Delta^2)/\sqrt{\pi\Delta^2}$ of width $\Delta$, we can then calculate the evolution of the one-body density matrix with the propagator (\ref{eq:evolcritical}). To meet the condition $\langle \hat{a}_{i}^{\dag}\hat{a}_{j}\rangle \simeq const$, the pulse duration $\tau$ must be small enough so that the interaction between atoms does not change the occupation distribution. This can be satisfied by operating the scattering pulse in the Raman-Nath regime~\cite{Denschlag2002JPB, Yue2013PRA}, \emph{i.e.}, $\tau\ll\hbar/\sqrt{U_{0}E_{R}}$ with $E_{R}=\hbar^{2}k^{2}/2m$ the single photon recoil energy. With this in mind, using further Eq.~(\ref{eq:momentum}), we get the momentum distribution before and after the scattering pulse, shown in Fig.~\ref{fig:fig2}. We choose the typical pulse experimental parameters as $U_{0}=80E_{R}$ and $\tau=3\mu s$, and typical density distribution width $\Delta\sim 100d$.

We see that when $\xi$ is very small (smaller than the thermal de Broglie wave length), the initial momentum distribution is of Gaussian type (Fig.~\ref{fig:fig2}(a)). After the KD scattering, the momentum distribution breaks into two peaks centered at $p_{z}\simeq\pm2\hbar k$. When the system is perfectly coherent like in a pure condensate, $g^{\left( 1\right)}=1$ over the whole system. In this case, the momentum distribution after the KD scattering is quite different. It has discrete peaks at $p_{z}=2n\hbar k$, $n=\pm1,\pm2,\cdots$(Fig.~\ref{fig:fig2}(d)).

It is worth pointing out that, in the calculation of the momentum distribution for the ultra-cold Bose gases in the critical regime, we have used the following form of the correlation function $g^{(1)}(r)$
\be
g^{(1)}(r) = \left\{ \begin{array}{l}\exp \left(-\pi r^2/\lambda_T^2\right), \quad \text{if}\quad r \le {r_c};\\[8pt]
r_0/r\exp\left(-r/\xi\right), \quad\text{if}\quad r > {r_c},
\end{array} \right.
\label{eq:corrfun}
\ee
with $r_c=\lambda _T\left(\lambda_T/\xi +\sqrt{\lambda_T^2/\xi^2+8\pi}\right)/4\pi$ and $r_0=r_c \exp\left(r_c/\xi-\pi r_c^2/\lambda_T^2 \right)$. In this form, $g^{(1)}(r)$ maintains a smooth function up to its first order derivative, besides the correct condition $g^{(1)}(0)=1$.

In the critical regime, when $\xi$ increases as temperature approaches the critical temperature, the final momentum distribution after the KD scattering has significant changes. Multi-peaks (Fig.~\ref{fig:fig2}(b) and (c)) at $p_{z}=2i\hbar k$, $i=\pm1,\pm2$ emerge from the two-peak momentum distribution of the thermal one. Further, as the correlation length increases, the peaks get sharper. To quantify the change of the momentum distribution with the correlation length, we define the visibility of the side peaks as
\begin{equation}
v=\frac{n_{2}+n_{4}-2n_{3}}{n_{2}+n_{4}+2n_{3}},
\label{eq:vdef}
\end{equation}
where $n_{j}$ is the momentum distribution at $p_{z}=j\hbar k$. Compared with the standard TOF method, measuring the visibility requires lower SNR, since we only need to measure the distribution at the specific momentum related to Eq.~(\ref{eq:vdef}), \emph{i.e.}, $j\hbar k$.

\begin{figure}[t]
\centering
\includegraphics[width=8.5cm]{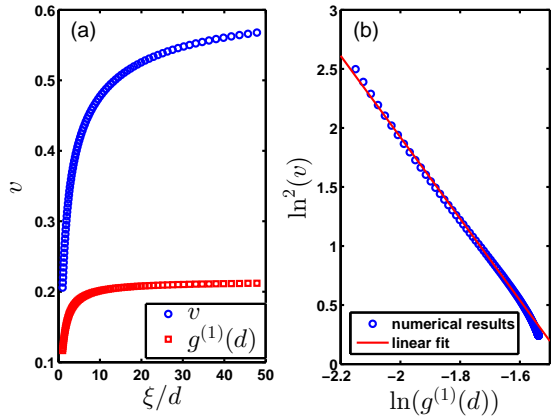}
\caption{(a) Visibility $v$ (blue circles) and correlation at $d$ (red squares) as a function of the correlation length $\xi$. (b) Dependence of $\ln^2(v)$ on $\ln(g^{(1)}(d)$. The blue circles are numerical results, and the red solid line is a linear fit.}
\label{fig:fig3}
\end{figure}

The dependence of the visibility defined above on the correlation length is then calculated as shown in Fig.~\ref{fig:fig3}(a). One sees that as the correlation length increases, the visibility increases monotonically. At this step, we arrive at the conclusion that the visibility can be used to identify the critical point, since the visibility maximizes at the critical temperature within the critical regime.

To show the physical picture of the dependence of the visibility on the correlation length, let us make use of the Raman-Nath approximation~\cite{Denschlag2002JPB, Yue2013PRA}, i.e., the duration of the KD scattering pulse is short enough. In this case, the propagator of the KD scattering process can be approximated by
\be
\begin{split}
K&\left( z,z^{\prime },t\right)  \simeq \left\langle z\right\vert \exp \left[ -i\frac{U_{0}t}{\hbar }\cos ^{2}\left( kz\right) \right] \left\vert z^{\prime }\right\rangle \\
&=e^{-i\Theta }\sum_{n}\left( -i\right) ^{n}J_{n}\left( \Theta \right) e^{2inkz} \delta \left( z-z^{\prime }\right) ,
\end{split}
\label{eq:KRN}
\ee
where $\Theta =U_{0}t/2\hbar $ and $J_{n}$ is the Bessel function of the first kind. In the derivation of the second line, we have used the Bessel expansion identities (Jacobi--Anger expansion): $\exp (iz\cos \theta )=\sum_{n}i^{n}J_{n}(z)\exp (in\theta)$. Substituting Eq. (\ref{eq:KRN}) into Eq. (\ref{eq:evolcritical}), it is straightforward to get the one-body density matrix at the end of the KD pulse
\be
\begin{split}
n^{\left( 1\right) }&\left( z_{1},z_{2},t\right)  =\frac{1}{\sqrt{\pi \Delta^{2}}}\sum_{n_{1},n_{2}}i^{n_{1}-n_{2}}J_{n_{1}}\left( \Theta \right)J_{n_{2}}\left( \Theta \right) \\
&\times e^{-2ik\left(n_{1}z_{1}-n_{2}z_{2}\right)}e^{-(z_{1}^{2}+z_{2}^{2})/2\Delta ^{2}}g^{\left( 1\right) }\left( z_{1}-z_{2}\right) .
\label{eq:Kobdm}
\end{split}
\ee
Let
$R=\left( z_{1}+z_{2}\right) /2,\ s=z_{1}-z_{2}$, and substituting Eq. (\ref{eq:Kobdm}) into Eq. (\ref{eq:momentum}), we get the corresponding momentum distribution,
\be
\begin{split}
n\left( p\right) =&\frac{1}{2\pi \hbar }\sum_{n_1,n_2} i^{n_1-n_2} J_{n_1} J_{n_2}e^{-k^2 \Delta
^2 \left( n_1-n_2 \right)^{2} }\\
&\times \int \mathrm{d}s e^{-s^2/4\Delta^2+i[p -\left( n_{1}+n_{2}\right) \hbar k] s/\hbar} g^{(1)}\left( s\right),
\end{split}
\ee
where we have omitted the parameter of the Bessel function for compactness. As $\Delta \gg d$, $k\Delta \gg 1$, $e^{ -k^{2}\Delta^{2}\left( n_{1}-n_{2}\right) ^{2}} \simeq \delta _{n_{1},n_{2}}$. Therefore the momentum distribution can be approximated as
\be
n\left( p\right) \simeq \frac{1}{2\pi \hbar }\sum_{n}J_{n}^{2}
\left( \Theta\right) I(p-2n\hbar k),
\label{eq:np1}
\ee
with
\be
I(p)=\int \mathrm{d}s\exp \left[ -\frac{s^{2}}{4\Delta ^{2}}+i
ps/\hbar \right] g^{\left( 1\right) }\left( s\right)
\ee
the Fourier transform of $\exp(-s^2/4\Delta^2)g^{(1)}(s)$.
Equation (\ref{eq:np1}) means that the momentum distribution after the pulse
is the summation of $I(p)$ shifted by $2n\hbar k$ with weight $J^2_n(\Theta)$.
As we know, the Fourier transform of the product of two functions equals the
convolution of their respective Fourier transform. Denote the Fourier transform
of $\exp(-s^2/4\Delta^2)$ and $g^{\left( 1\right) }\left( s\right)$ by
$F\left( p\right) $ and $G\left( p\right) $, respectively,
\begin{equation}
I(p)=F(p)\otimes G(p),
\end{equation}
where $\otimes$ denotes convolution. Notice that
$
F\left( p\right) =2\sqrt{\pi }\Delta \exp \left( -p^{2}\Delta ^{2}/\hbar
^{2}\right).
$
When $\Delta \gg d$, $F\left( p\right)\simeq\delta \left(p\right) $. Therefore
we have $I\left( p\right) \simeq G\left( p\right) $, and
\begin{equation}
n\left( p\right) \simeq \frac{1}{2\pi \hbar }\sum_{n}J_{n}^{2}\left( \Theta
\right) G\left( p+2n\hbar k\right) ,
\label{eq:npf}
\end{equation}
which means that the final momentum distribution is the summation of the
Fourier transform of the correlation function $g^{(1)}(s)$ shifted by $2n\hbar
k$ with weight $J_{n}^{2}\left( \Theta \right)$. It is pointed out that Equation (\ref{eq:npf}) is only an approximate result under the Raman-Nath (RN) approximation, where the kinetic energy induced phases are entirely ignored~\cite{Yue2013PRA}. Besides, we assume the initial wave package width much larger than the lattice constant, which may be not the case when the atomic number is not so large or the confinement by the trap is strong. Nevertheless, We can use it to estimate the visibility (\ref{eq:vdef}) in a qualitative way.

Within the approximate momentum distribution (\ref{eq:npf}), the visibility is expressed as
\begin{widetext}
\begin{equation}
v = \frac{\sum_{n}J_{n}^{2}\left( \Theta \right) \left[ G\left( \left(2n+2\right) \hbar k\right) +G\left( \left( 2n+4\right) \hbar k\right)-2G\left( \left( 2n+3\right) \hbar k\right) \right] }{\sum_{n}J_{n}^{2}\left( \Theta \right) \left[ G\left( \left( 2n+2\right) \hbar k\right)+G\left( \left( 2n+4\right) \hbar k\right) +2G\left( \left( 2n+3\right)\hbar k\right) \right] }.
\end{equation}
With further approximation that the weight $J^2_n(\Theta)$ of different $n$ is equal, the visibility becomes
\be
v=\frac{\sum_{n}G\left( 2n\hbar k\right) -G\left( \left( 2n+1\right)\hbar k\right) }{\sum_{n} G\left( 2n\hbar k\right) +G\left(\left( 2n+1\right) \hbar k\right) }=\frac{\int \mathrm{d}sg^{\left( 1\right) }\left( s\right) \sum_{n}\left(-1\right) ^{n}e^{inks}}{\int \mathrm{d}sg^{\left( 1\right) }\left( s\right)\sum_{n}e^{inks}}.
\ee
\end{widetext}
The $(-1)^n$ in the numerator at the right hand side comes from the alternate sign of the odd and even terms in the momentum space.
As $\lim_{n\rightarrow\infty}\sum_{n}\left( -1\right)^{n}e^{inks}=\sum_{n}\delta \left( s-\left( 2n+1\right) d\right) $ and $\lim_{n\rightarrow\infty}\sum_{n}e^{inks}=\sum_{n}\delta \left( s-2nd\right) $, it is easy to get
\begin{equation}
\begin{split}
v&=\frac{\sum_{n}g^{\left( 1\right) }\left( \left( 2n+1\right) d\right) }{\sum_{n}g^{\left( 1\right) }\left( 2nd\right) }.
\end{split}
\label{eq:vgd}
\end{equation}
By making the lowest-order approximation, we obtain
\be
\begin{split}
v\simeq &\frac{2g^{\left(1\right) }\left( d\right) }{1+2g^{\left( 1\right) }\left( 2d\right) }\\
=&\frac{2g^{\left( 1\right) }\left( d\right) }{1+d/r_{0}\left( g^{\left( 1\right)}\left( d\right) \right) ^{2}},
\end{split}
\label{eq:vgd1}
\ee
where $g^{\left( 1\right) }\left( d\right) $ is the correlation at distance of lattice constant $d$, another length scale of the system.

Equation (\ref{eq:vgd}) and (\ref{eq:vgd1}) show that with the approximation that $J_{n}\left( \Theta \right)$ is equal for different $n$, the visibility is a simple function of the correlation at $d$. When the weight $J_{n}^2\left( \Theta \right) $ of different $n$ is different, only finite orders contribute significantly. With the parameters we consider for the KD scattering in this work, we know that $J_{0}\simeq 0, J_{\pm 1}\sim J_{\pm 2}\gg J_{\left\vert n\right\vert >2}$. It is hard for us to get a simple relation between $v$ and $g^{(1)}(d)$ in this situation, but we still anticipate that they are simply related because of the length scale $d$ introduced by the KD scattering.

The above discussions give us the idea to consider the relation between $v$ and $g^{(1)}(d)$. In Fig.\ref{fig:fig3}(a), we plot both $v$ and $g^{\left(1\right) }\left( d\right) $ as a function of $\xi $. It is quite interesting to find that $\ln^{2}(v)$ depends on $\ln \left( g^{\left( 1\right) }\left( d\right)\right) $ linearly in a considerable range (Fig.\ref{fig:fig3}(b)). The relation can be written as
\begin{equation}
\ln ^{2}(v)=-c_1\ln \left( g^{\left( 1\right) }\left( d\right) \right) +c_2\,,
\end{equation}
with $-c_1$ and $c_2$ the slope and the interception of the line, respectively. Notice that $c_1$ and $c_2$ depend on each other. The relation between $c_1$ and $c_2$ can be found by considering a limit case. As $\xi \rightarrow \infty $, the visibility $v\rightarrow 1$ and $\ln (v)\rightarrow 0$, so the linear fitting line intersects the horizontal axis at $g^{(1)}(d)|_{\xi =\infty }$, i.e.
\begin{equation}
c_2=c_1\ln\left(g^{\left(1\right)} \left(d\right)|_{\xi=\infty}\right).
\end{equation}
Note also that the slope $c_1$ is a systematic parameter which is determined by the dependence of the visibility $v$ and the correlation $g^{(1)}$ on the correlation length $\xi$.

In the critical regime that we are discussing, $g^{\left(1\right)}\left(d\right)=r_0e^{-d/\xi}/d$. In this case, one gets
\begin{equation}
c_2=c_1\ln\left(\frac{r_0}{d}\right).
\end{equation}
In a straightforward manner, we obtain
\begin{equation}
\ln(v)=-\sqrt{\frac{c_1d}{\xi}}.
\end{equation}
Here we take the minus root since $v\le1$.

As we have mentioned, the correlation length $\xi=\xi_0\left\vert t\right\vert ^{-\nu}$ with $\xi_0$ the amplitude of the correlation length and $t=1-T/T_{c}$ the relative temperature. Then we have
\begin{equation}
v(t)=\exp\left(-a_v\left\vert t\right\vert ^{\nu/2}\right)\,,
\label{eq:fv}
\end{equation}
with
\begin{equation}
a_v=\sqrt{\frac{c_1d}{\xi_0}}.
\end{equation}
Equation~(\ref{eq:fv}) shows that we can extract the critical exponent by measuring the visibility at different temperature. Furthermore, from the measurement above and below the critical temperature, we can also get the correlation amplitude ratio $\xi_0^>/\xi_0^<=(a_v^</a_v^>)^2$ from the fitting to get $a_v$ above and below the critical temperature.

\begin{figure}[t]
\centering
\includegraphics[width=8cm]{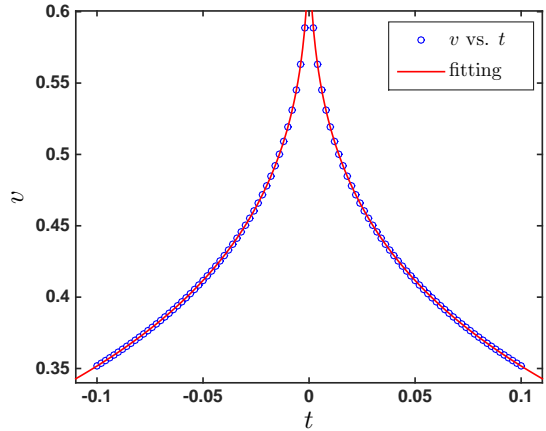}
\caption{Numerical results of the dependence of the visibility $v$ on the relative temperature $t$ in the critical regime. The solid lines are fitting results according to Eq. (\protect\ref{eq:fv}).}
\label{fig:fig4}
\end{figure}

To see how precise Eq.~(\ref{eq:fv}) can be, we use it to fit the numerical results. We first get the correlation length by $\xi=\xi_0 |t|^\nu$ with the standard theoretical value of the critical exponent $\nu=0.67$ at different relative temperature. Then we calculate the correlation function according to Eq.~(\ref{eq:corrfun}). With the propagator (\ref{eq:propogator}) and the initial density distribution width $\Delta$, we get the one-body density matrix at the end of the pulse and the momentum distribution according to Eq.~(\ref{eq:evolcritical}) and Eq.~(\ref{eq:momentum}), respectively. The visibility $v$ at different relative temperature is obtained by the definition (\ref{eq:vdef}), the result of which are shown by the blue circles in Fig.~\ref{fig:fig4}.
We fit them by Eq.~(\ref{eq:fv}) with $\nu$ and $a_v$ the free fitting parameters. The fitting result is shown by the red solid lines in Fig.~\ref{fig:fig4}. The critical exponent obtained by the fitting is
\begin{equation}
\nu=0.67\pm0.01,
\end{equation}
which is consistent with what we have assumed in our calculation. The results shows that we can indeed extract the critical exponent with Eq.~(\ref{eq:fv}).

\section{Summary and Discussion}
\label{Sec:Disc}

In summary, the KD scattering is proposed to observe the critical phase transition of ultra-cold Bose atomic gases. Our theory shows that this method has the potential to be a powerful way to measure the critical behavior such as the critical exponent and correlation amplitude ratio. In particular, our theoretical simulation shows that a very simple relation can be used to extract the critical exponent and correlation amplitude ratio from the fitting of the measured visibility for different temperature.

During the early stage of the free expansion, the interaction energy will be converted to kinetic energy, and the momentum distribution is usually changed~\cite{Stenger1999PRL}. However, what we consider is a quasi-one-dimensional system, which expands very fast along the initially strongly confined directions. Owing to this fast density drop, the inter-particle interactions play nearly no role during the TOF, and the expansion in the $z$ direction is governed essentially by the initial momentum distribution before the expansion. We can further reduce the interaction effect by operating at small particle number.

Compared with standard TOF technique, the present method makes use of the large correlation length in the critical regime to produce interference fringes. Furthermore, the measurement of the visibility of the interference fringes demands lower signal to noise ratio than measuring full momentum distribution in standard TOF Technique. Compared with the Talbot-Lau interferometer~\cite{Xiong2013LPL} to study the critical behavior, the present method has the merit that the theory is simpler. In particular, a simple relation is found to extract the critical behavior. The experimental advances of the ultra-cold atomic gases confined in a box potential~\cite{Gaunt2013PRL, Gotlib2014PRA, Chomaz2015nc} make it very promising to study the critical behavior more precisely with KD scattering. In addition, we believe KD scattering has the potential to study other novel ultra-cold atomic gases such as fermionic atomic gases.

\section*{Acknowledgement}
X.Y. thanks helpful discussion with Bo Liu and Biao Wu. This work was supported by National Key Basic Research and Development Program of China under Grant No. 2011CB921503 and NSFC 11175246, 11334001, and 11504328.

\bibliography{Correlation-KDscattering}

\begin{thebibliography}{25}%
\makeatletter
\providecommand \@ifxundefined [1]{%
 \@ifx{#1\undefined}
}%
\providecommand \@ifnum [1]{%
 \ifnum #1\expandafter \@firstoftwo
 \else \expandafter \@secondoftwo
 \fi
}%
\providecommand \@ifx [1]{%
 \ifx #1\expandafter \@firstoftwo
 \else \expandafter \@secondoftwo
 \fi
}%
\providecommand \natexlab [1]{#1}%
\providecommand \enquote  [1]{``#1''}%
\providecommand \bibnamefont  [1]{#1}%
\providecommand \bibfnamefont [1]{#1}%
\providecommand \citenamefont [1]{#1}%
\providecommand \href@noop [0]{\@secondoftwo}%
\providecommand \href [0]{\begingroup \@sanitize@url \@href}%
\providecommand \@href[1]{\@@startlink{#1}\@@href}%
\providecommand \@@href[1]{\endgroup#1\@@endlink}%
\providecommand \@sanitize@url [0]{\catcode `\\12\catcode `\$12\catcode
  `\&12\catcode `\#12\catcode `\^12\catcode `\_12\catcode `\%12\relax}%
\providecommand \@@startlink[1]{}%
\providecommand \@@endlink[0]{}%
\providecommand \url  [0]{\begingroup\@sanitize@url \@url }%
\providecommand \@url [1]{\endgroup\@href {#1}{\urlprefix }}%
\providecommand \urlprefix  [0]{URL }%
\providecommand \Eprint [0]{\href }%
\providecommand \doibase [0]{http://dx.doi.org/}%
\providecommand \selectlanguage [0]{\@gobble}%
\providecommand \bibinfo  [0]{\@secondoftwo}%
\providecommand \bibfield  [0]{\@secondoftwo}%
\providecommand \translation [1]{[#1]}%
\providecommand \BibitemOpen [0]{}%
\providecommand \bibitemStop [0]{}%
\providecommand \bibitemNoStop [0]{.\EOS\space}%
\providecommand \EOS [0]{\spacefactor3000\relax}%
\providecommand \BibitemShut  [1]{\csname bibitem#1\endcsname}%
\let\auto@bib@innerbib\@empty
\bibitem [{\citenamefont {Zinn-Justin}(2013)}]{Zinn-Justin2013}%
  \BibitemOpen
  \bibfield  {author} {\bibinfo {author} {\bibfnamefont {J.}~\bibnamefont
  {Zinn-Justin}},\ }\href@noop {} {\emph {\bibinfo {title} {Phase Transitions
  and Renormalization Group}}}\ (\bibinfo  {publisher} {Oxford University
  Press},\ \bibinfo {year} {2013})\BibitemShut {NoStop}%
\bibitem [{\citenamefont {Navon}\ \emph {et~al.}(2015)\citenamefont {Navon},
  \citenamefont {Gaunt}, \citenamefont {Smith},\ and\ \citenamefont
  {Hadzibabic}}]{Navon2015Sci}%
  \BibitemOpen
  \bibfield  {author} {\bibinfo {author} {\bibfnamefont {N.}~\bibnamefont
  {Navon}}, \bibinfo {author} {\bibfnamefont {A.~L.}\ \bibnamefont {Gaunt}},
  \bibinfo {author} {\bibfnamefont {R.~P.}\ \bibnamefont {Smith}}, \ and\
  \bibinfo {author} {\bibfnamefont {Z.}~\bibnamefont {Hadzibabic}},\
  }\href@noop {} {\bibfield  {journal} {\bibinfo  {journal} {Science}\ }\textbf
  {\bibinfo {volume} {347}},\ \bibinfo {pages} {167} (\bibinfo {year}
  {2015})}\BibitemShut {NoStop}%
\bibitem [{\citenamefont {Zurek}(1996)}]{Zurek1996PhysRep}%
  \BibitemOpen
  \bibfield  {author} {\bibinfo {author} {\bibfnamefont {W.~H.}\ \bibnamefont
  {Zurek}},\ }\href@noop {} {\bibfield  {journal} {\bibinfo  {journal} {Phys.
  Rep.}\ }\textbf {\bibinfo {volume} {276}},\ \bibinfo {pages} {177} (\bibinfo
  {year} {1996})}\BibitemShut {NoStop}%
\bibitem [{\citenamefont {Sachdev}(2011)}]{Sachdev2011}%
  \BibitemOpen
  \bibfield  {author} {\bibinfo {author} {\bibfnamefont {S.}~\bibnamefont
  {Sachdev}},\ }\href@noop {} {\emph {\bibinfo {title} {Quantum Phase
  Transitions}}}\ (\bibinfo  {publisher} {Cambridge University Press},\
  \bibinfo {year} {2011})\BibitemShut {NoStop}%
\bibitem [{\citenamefont {Zinn-Justin}(2002)}]{Zinn-Justin2002}%
  \BibitemOpen
  \bibfield  {author} {\bibinfo {author} {\bibfnamefont {J.}~\bibnamefont
  {Zinn-Justin}},\ }\href@noop {} {\emph {\bibinfo {title} {Quantum Field
  Theory and Critical Phenomena}}}\ (\bibinfo  {publisher} {Oxford University
  Press},\ \bibinfo {year} {2002})\BibitemShut {NoStop}%
\bibitem [{\citenamefont {Bloch}\ \emph {et~al.}(2008)\citenamefont {Bloch},
  \citenamefont {Dalibard},\ and\ \citenamefont {Zwerger}}]{Bloch2008RMP}%
  \BibitemOpen
  \bibfield  {author} {\bibinfo {author} {\bibfnamefont {I.}~\bibnamefont
  {Bloch}}, \bibinfo {author} {\bibfnamefont {J.}~\bibnamefont {Dalibard}}, \
  and\ \bibinfo {author} {\bibfnamefont {W.}~\bibnamefont {Zwerger}},\
  }\href@noop {} {\bibfield  {journal} {\bibinfo  {journal} {Rev. Mod. Phys.}\
  }\textbf {\bibinfo {volume} {80}},\ \bibinfo {pages} {885} (\bibinfo {year}
  {2008})}\BibitemShut {NoStop}%
\bibitem [{\citenamefont {Stoof}\ \emph {et~al.}(2008)\citenamefont {Stoof},
  \citenamefont {Gubbels},\ and\ \citenamefont {Dickerscheid}}]{Stoof2008}%
  \BibitemOpen
  \bibfield  {author} {\bibinfo {author} {\bibfnamefont {H.}~\bibnamefont
  {Stoof}}, \bibinfo {author} {\bibfnamefont {K.}~\bibnamefont {Gubbels}}, \
  and\ \bibinfo {author} {\bibfnamefont {D.}~\bibnamefont {Dickerscheid}},\
  }\href@noop {} {\emph {\bibinfo {title} {Ultracold Quantum Fields}}}\
  (\bibinfo  {publisher} {Springer Netherlands},\ \bibinfo {year}
  {2008})\BibitemShut {NoStop}%
\bibitem [{\citenamefont {Donner}\ \emph {et~al.}(2007)\citenamefont {Donner},
  \citenamefont {Ritter}, \citenamefont {Bourdel}, \citenamefont {\"{O}ttl},
  \citenamefont {K\"ohl},\ and\ \citenamefont {Esslinger}}]{Donner2007Science}%
  \BibitemOpen
  \bibfield  {author} {\bibinfo {author} {\bibfnamefont {T.}~\bibnamefont
  {Donner}}, \bibinfo {author} {\bibfnamefont {S.}~\bibnamefont {Ritter}},
  \bibinfo {author} {\bibfnamefont {T.}~\bibnamefont {Bourdel}}, \bibinfo
  {author} {\bibfnamefont {A.}~\bibnamefont {\"{O}ttl}}, \bibinfo {author}
  {\bibfnamefont {M.}~\bibnamefont {K\"ohl}}, \ and\ \bibinfo {author}
  {\bibfnamefont {T.}~\bibnamefont {Esslinger}},\ }\href@noop {} {\bibfield
  {journal} {\bibinfo  {journal} {Science}\ }\textbf {\bibinfo {volume}
  {315}},\ \bibinfo {pages} {1556} (\bibinfo {year} {2007})}\BibitemShut
  {NoStop}%
\bibitem [{\citenamefont {Xiong}\ \emph {et~al.}(2013)\citenamefont {Xiong},
  \citenamefont {Zhou}, \citenamefont {Yue}, \citenamefont {Chen},
  \citenamefont {Wu},\ and\ \citenamefont {Xiong}}]{Xiong2013LPL}%
  \BibitemOpen
  \bibfield  {author} {\bibinfo {author} {\bibfnamefont {W.}~\bibnamefont
  {Xiong}}, \bibinfo {author} {\bibfnamefont {X.}~\bibnamefont {Zhou}},
  \bibinfo {author} {\bibfnamefont {X.}~\bibnamefont {Yue}}, \bibinfo {author}
  {\bibfnamefont {X.}~\bibnamefont {Chen}}, \bibinfo {author} {\bibfnamefont
  {B.}~\bibnamefont {Wu}}, \ and\ \bibinfo {author} {\bibfnamefont
  {H.}~\bibnamefont {Xiong}},\ }\href@noop {} {\bibfield  {journal} {\bibinfo
  {journal} {Laser Phys. Lett.}\ }\textbf {\bibinfo {volume} {10}},\ \bibinfo
  {pages} {125502} (\bibinfo {year} {2013})}\BibitemShut {NoStop}%
\bibitem [{\citenamefont {Chomaz}\ \emph {et~al.}(2015)\citenamefont {Chomaz},
  \citenamefont {Corman}, \citenamefont {Bienaim\'{e}}, \citenamefont
  {Desbuquois}, \citenamefont {Weitenberg}, \citenamefont {Nascimb\`{e}ne},
  \citenamefont {Beugnon},\ and\ \citenamefont {Dalibard}}]{Chomaz2015nc}%
  \BibitemOpen
  \bibfield  {author} {\bibinfo {author} {\bibfnamefont {L.}~\bibnamefont
  {Chomaz}}, \bibinfo {author} {\bibfnamefont {L.}~\bibnamefont {Corman}},
  \bibinfo {author} {\bibfnamefont {T.}~\bibnamefont {Bienaim\'{e}}}, \bibinfo
  {author} {\bibfnamefont {R.}~\bibnamefont {Desbuquois}}, \bibinfo {author}
  {\bibfnamefont {C.}~\bibnamefont {Weitenberg}}, \bibinfo {author}
  {\bibfnamefont {S.}~\bibnamefont {Nascimb\`{e}ne}}, \bibinfo {author}
  {\bibfnamefont {J.}~\bibnamefont {Beugnon}}, \ and\ \bibinfo {author}
  {\bibfnamefont {J.}~\bibnamefont {Dalibard}},\ }\href@noop {} {\bibfield
  {journal} {\bibinfo  {journal} {Nat. Commun.}\ }\textbf {\bibinfo {volume}
  {6}},\ \bibinfo {pages} {6162} (\bibinfo {year} {2015})}\BibitemShut
  {NoStop}%
\bibitem [{\citenamefont {Cronin}\ \emph {et~al.}(2009)\citenamefont {Cronin},
  \citenamefont {Schmiedmayer},\ and\ \citenamefont
  {Pritchard}}]{Cronin2009RMP}%
  \BibitemOpen
  \bibfield  {author} {\bibinfo {author} {\bibfnamefont {A.~D.}\ \bibnamefont
  {Cronin}}, \bibinfo {author} {\bibfnamefont {J.}~\bibnamefont
  {Schmiedmayer}}, \ and\ \bibinfo {author} {\bibfnamefont {D.~E.}\
  \bibnamefont {Pritchard}},\ }\href@noop {} {\bibfield  {journal} {\bibinfo
  {journal} {Rev. Mod. Phys.}\ }\textbf {\bibinfo {volume} {81}},\ \bibinfo
  {pages} {1051} (\bibinfo {year} {2009})}\BibitemShut {NoStop}%
\bibitem [{\citenamefont {Yue}\ \emph {et~al.}(2013)\citenamefont {Yue},
  \citenamefont {Zhai}, \citenamefont {Wang}, \citenamefont {Xiong},
  \citenamefont {Chen},\ and\ \citenamefont {Zhou}}]{Yue2013PRA}%
  \BibitemOpen
  \bibfield  {author} {\bibinfo {author} {\bibfnamefont {X.}~\bibnamefont
  {Yue}}, \bibinfo {author} {\bibfnamefont {Y.}~\bibnamefont {Zhai}}, \bibinfo
  {author} {\bibfnamefont {Z.}~\bibnamefont {Wang}}, \bibinfo {author}
  {\bibfnamefont {H.}~\bibnamefont {Xiong}}, \bibinfo {author} {\bibfnamefont
  {X.}~\bibnamefont {Chen}}, \ and\ \bibinfo {author} {\bibfnamefont
  {X.}~\bibnamefont {Zhou}},\ }\href@noop {} {\bibfield  {journal} {\bibinfo
  {journal} {Phys. Rev. A}\ }\textbf {\bibinfo {volume} {88}},\ \bibinfo
  {pages} {013603} (\bibinfo {year} {2013})}\BibitemShut {NoStop}%
\bibitem [{\citenamefont {Polkovnikov}\ \emph {et~al.}(2011)\citenamefont
  {Polkovnikov}, \citenamefont {Sengupta}, \citenamefont {Silva},\ and\
  \citenamefont {Vengalattore}}]{Polkovnikov2011RMP}%
  \BibitemOpen
  \bibfield  {author} {\bibinfo {author} {\bibfnamefont {A.}~\bibnamefont
  {Polkovnikov}}, \bibinfo {author} {\bibfnamefont {K.}~\bibnamefont
  {Sengupta}}, \bibinfo {author} {\bibfnamefont {A.}~\bibnamefont {Silva}}, \
  and\ \bibinfo {author} {\bibfnamefont {M.}~\bibnamefont {Vengalattore}},\
  }\href@noop {} {\bibfield  {journal} {\bibinfo  {journal} {Rev. Mod. Phys.}\
  }\textbf {\bibinfo {volume} {83}},\ \bibinfo {pages} {863} (\bibinfo {year}
  {2011})}\BibitemShut {NoStop}%
\bibitem [{\citenamefont {Penrose}\ and\ \citenamefont
  {Onsager}(1956)}]{Penrose1956PR}%
  \BibitemOpen
  \bibfield  {author} {\bibinfo {author} {\bibfnamefont {O.}~\bibnamefont
  {Penrose}}\ and\ \bibinfo {author} {\bibfnamefont {L.}~\bibnamefont
  {Onsager}},\ }\href@noop {} {\bibfield  {journal} {\bibinfo  {journal} {Phys.
  Rev.}\ }\textbf {\bibinfo {volume} {104}},\ \bibinfo {pages} {576} (\bibinfo
  {year} {1956})}\BibitemShut {NoStop}%
\bibitem [{\citenamefont {Pitaevskii}\ and\ \citenamefont
  {Stringari}()}]{Pitaevskii2003}%
  \BibitemOpen
  \bibfield  {author} {\bibinfo {author} {\bibfnamefont {L.~P.}\ \bibnamefont
  {Pitaevskii}}\ and\ \bibinfo {author} {\bibfnamefont {S.}~\bibnamefont
  {Stringari}},\ }\href@noop {} {\emph {\bibinfo {title} {Bose-Einstein
  Condensation}}},\ International Series of Monographs on Physics\ (\bibinfo
  {publisher} {Oxford University Press})\BibitemShut {NoStop}%
\bibitem [{\citenamefont {Ueda}(2010)}]{Ueda2010}%
  \BibitemOpen
  \bibfield  {author} {\bibinfo {author} {\bibfnamefont {M.}~\bibnamefont
  {Ueda}},\ }\href@noop {} {\emph {\bibinfo {title} {Fundamentals and New
  Frontiers of Bose-Einstein Condensation}}}\ (\bibinfo  {publisher} {World
  Scientific},\ \bibinfo {year} {2010})\BibitemShut {NoStop}%
\bibitem [{\citenamefont {Wilson}(1975)}]{Wilson1975RMP}%
  \BibitemOpen
  \bibfield  {author} {\bibinfo {author} {\bibfnamefont {K.~G.}\ \bibnamefont
  {Wilson}},\ }\href@noop {} {\bibfield  {journal} {\bibinfo  {journal} {Rev.
  Mod. Phys.}\ }\textbf {\bibinfo {volume} {47}},\ \bibinfo {pages} {773}
  (\bibinfo {year} {1975})}\BibitemShut {NoStop}%
\bibitem [{\citenamefont {Huang}(1987)}]{Huang1987}%
  \BibitemOpen
  \bibfield  {author} {\bibinfo {author} {\bibfnamefont {K.}~\bibnamefont
  {Huang}},\ }\href@noop {} {\emph {\bibinfo {title} {Statistical
  Mechanics}}},\ \bibinfo {edition} {2nd}\ ed.\ (\bibinfo  {publisher} {Wiley
  India Pvt. Limited},\ \bibinfo {year} {1987})\BibitemShut {NoStop}%
\bibitem [{\citenamefont {Privman}\ \emph {et~al.}(1991)\citenamefont
  {Privman}, \citenamefont {Hohenberg},\ and\ \citenamefont
  {Aharony}}]{Privman1991}%
  \BibitemOpen
  \bibfield  {author} {\bibinfo {author} {\bibfnamefont {V.}~\bibnamefont
  {Privman}}, \bibinfo {author} {\bibfnamefont {P.~C.}\ \bibnamefont
  {Hohenberg}}, \ and\ \bibinfo {author} {\bibfnamefont {A.}~\bibnamefont
  {Aharony}},\ }in\ \href@noop {} {\emph {\bibinfo {booktitle} {Phase
  transitions and critical phenomena}}},\ Vol.~\bibinfo {volume} {14},\
  \bibinfo {editor} {edited by\ \bibinfo {editor} {\bibfnamefont
  {C.}~\bibnamefont {Domb}}\ and\ \bibinfo {editor} {\bibfnamefont {J.~L.}\
  \bibnamefont {Lebowitz}}}\ (\bibinfo  {publisher} {Academic Press},\ \bibinfo
  {year} {1991})\ pp.\ \bibinfo {pages} {1--134}\BibitemShut {NoStop}%
\bibitem [{\citenamefont {Chaikin}\ and\ \citenamefont
  {Lubensky}(2000)}]{Chaikin2000}%
  \BibitemOpen
  \bibfield  {author} {\bibinfo {author} {\bibfnamefont {P.~M.}\ \bibnamefont
  {Chaikin}}\ and\ \bibinfo {author} {\bibfnamefont {T.~C.}\ \bibnamefont
  {Lubensky}},\ }\href@noop {} {\emph {\bibinfo {title} {Principles of
  Condensed Matter Physics}}}\ (\bibinfo  {publisher} {Cambridge University
  Press},\ \bibinfo {year} {2000})\BibitemShut {NoStop}%
\bibitem [{\citenamefont {Denschlag}\ \emph {et~al.}(2002)\citenamefont
  {Denschlag}, \citenamefont {Simsarian}, \citenamefont {H{\"a}ffner},
  \citenamefont {McKenzie}, \citenamefont {Browaeys}, \citenamefont {Cho},
  \citenamefont {Helmerson}, \citenamefont {Rolston},\ and\ \citenamefont
  {Phillips}}]{Denschlag2002JPB}%
  \BibitemOpen
  \bibfield  {author} {\bibinfo {author} {\bibfnamefont {J.~H.}\ \bibnamefont
  {Denschlag}}, \bibinfo {author} {\bibfnamefont {J.~E.}\ \bibnamefont
  {Simsarian}}, \bibinfo {author} {\bibfnamefont {H.}~\bibnamefont
  {H{\"a}ffner}}, \bibinfo {author} {\bibfnamefont {C.}~\bibnamefont
  {McKenzie}}, \bibinfo {author} {\bibfnamefont {A.}~\bibnamefont {Browaeys}},
  \bibinfo {author} {\bibfnamefont {D.}~\bibnamefont {Cho}}, \bibinfo {author}
  {\bibfnamefont {K.}~\bibnamefont {Helmerson}}, \bibinfo {author}
  {\bibfnamefont {S.~L.}\ \bibnamefont {Rolston}}, \ and\ \bibinfo {author}
  {\bibfnamefont {W.~D.}\ \bibnamefont {Phillips}},\ }\href@noop {} {\bibfield
  {journal} {\bibinfo  {journal} {J. Phys. B}\ }\textbf {\bibinfo {volume}
  {35}},\ \bibinfo {pages} {3095} (\bibinfo {year} {2002})}\BibitemShut
  {NoStop}%
\bibitem [{\citenamefont {Morsch}\ and\ \citenamefont
  {Oberthaler}(2006)}]{Morsch2006RMP}%
  \BibitemOpen
  \bibfield  {author} {\bibinfo {author} {\bibfnamefont {O.}~\bibnamefont
  {Morsch}}\ and\ \bibinfo {author} {\bibfnamefont {M.}~\bibnamefont
  {Oberthaler}},\ }\href@noop {} {\bibfield  {journal} {\bibinfo  {journal}
  {Rev. Mod. Phys.}\ }\textbf {\bibinfo {volume} {78}},\ \bibinfo {pages} {179}
  (\bibinfo {year} {2006})}\BibitemShut {NoStop}%
\bibitem [{\citenamefont {Stenger}\ \emph {et~al.}(1999)\citenamefont
  {Stenger}, \citenamefont {Inouye}, \citenamefont {Chikkatur}, \citenamefont
  {Stamper-Kurn}, \citenamefont {Pritchard},\ and\ \citenamefont
  {Ketterle}}]{Stenger1999PRL}%
  \BibitemOpen
  \bibfield  {author} {\bibinfo {author} {\bibfnamefont {J.}~\bibnamefont
  {Stenger}}, \bibinfo {author} {\bibfnamefont {S.}~\bibnamefont {Inouye}},
  \bibinfo {author} {\bibfnamefont {A.}~\bibnamefont {Chikkatur}}, \bibinfo
  {author} {\bibfnamefont {D.}~\bibnamefont {Stamper-Kurn}}, \bibinfo {author}
  {\bibfnamefont {D.}~\bibnamefont {Pritchard}}, \ and\ \bibinfo {author}
  {\bibfnamefont {W.}~\bibnamefont {Ketterle}},\ }\href@noop {} {\bibfield
  {journal} {\bibinfo  {journal} {Phys. Rev. Lett.}\ }\textbf {\bibinfo
  {volume} {82}},\ \bibinfo {pages} {4569} (\bibinfo {year}
  {1999})}\BibitemShut {NoStop}%
\bibitem [{\citenamefont {Gaunt}\ \emph {et~al.}(2013)\citenamefont {Gaunt},
  \citenamefont {Schmidutz}, \citenamefont {Gotlibovych}, \citenamefont
  {Smith},\ and\ \citenamefont {Hadzibabic}}]{Gaunt2013PRL}%
  \BibitemOpen
  \bibfield  {author} {\bibinfo {author} {\bibfnamefont {A.~L.}\ \bibnamefont
  {Gaunt}}, \bibinfo {author} {\bibfnamefont {T.~F.}\ \bibnamefont
  {Schmidutz}}, \bibinfo {author} {\bibfnamefont {I.}~\bibnamefont
  {Gotlibovych}}, \bibinfo {author} {\bibfnamefont {R.~P.}\ \bibnamefont
  {Smith}}, \ and\ \bibinfo {author} {\bibfnamefont {Z.}~\bibnamefont
  {Hadzibabic}},\ }\href@noop {} {\bibfield  {journal} {\bibinfo  {journal}
  {Phys. Rev. Lett.}\ }\textbf {\bibinfo {volume} {110}},\ \bibinfo {pages}
  {200406} (\bibinfo {year} {2013})}\BibitemShut {NoStop}%
\bibitem [{\citenamefont {Gotlibovych}\ \emph {et~al.}(2014)\citenamefont
  {Gotlibovych}, \citenamefont {Schmidutz}, \citenamefont {Gaunt},
  \citenamefont {Navon}, \citenamefont {Smith},\ and\ \citenamefont
  {Hadzibabic}}]{Gotlib2014PRA}%
  \BibitemOpen
  \bibfield  {author} {\bibinfo {author} {\bibfnamefont {I.}~\bibnamefont
  {Gotlibovych}}, \bibinfo {author} {\bibfnamefont {T.~F.}\ \bibnamefont
  {Schmidutz}}, \bibinfo {author} {\bibfnamefont {A.~L.}\ \bibnamefont
  {Gaunt}}, \bibinfo {author} {\bibfnamefont {N.}~\bibnamefont {Navon}},
  \bibinfo {author} {\bibfnamefont {R.~P.}\ \bibnamefont {Smith}}, \ and\
  \bibinfo {author} {\bibfnamefont {Z.}~\bibnamefont {Hadzibabic}},\
  }\href@noop {} {\bibfield  {journal} {\bibinfo  {journal} {Phys. Rev. A}\
  }\textbf {\bibinfo {volume} {89}},\ \bibinfo {pages} {061604} (\bibinfo
  {year} {2014})}\BibitemShut {NoStop}%
\end{thebibliography}%
\end{document}